# Application of Local Approaches to the Assessment of Fatigue Test results obtained for Welded Joints at Sub-Zero Temperatures

Moritz Braun[a], Aleksandar-Saša Milaković[a], Finn Renken[a], Wolfgang Fricke[a], Sören Ehlers[a]

[a] *Institute for Ship Structural Design and Analysis, Hamburg University of Technology, Hamburg, Germany*

Corresponding author:

Moritz Braun (moritz.br@tuhh.de)

Abstract:

Several studies have found significant increase in the fatigue strength of welded joints of structural steels at sub-zero temperatures. This study addresses the research by investigating the applicability of local fatigue assessment methods to welded joints exposed to sub-zero temperatures. For this purpose, fatigue test results of two fillet weld details with weld toe and weld root failure are evaluated at a range of temperatures using a variety of structural hot-spot and notch stress approaches, then are compared to the nominal stress approach. Large differences in prediction accuracy are found for the analysed assessment methods and both failure locations.

1. **Introduction**

Presently, many different local fatigue assessment procedures have been proposed for welded structures. Most fatigue assessment methods used in structural design, such as the nominal, the structural hot-spot or notch stress approach, were originally developed for structures without considering temperature variations. Moreover, since fatigue tests are usually performed in laboratories at room temperature (RT), fatigue test data and methods are usually only valid for a temperature around 20 °C. Thus, temperature effects are usually neglected within fatigue assessment—with the exception of very high temperatures, where significant softening of structural materials or creeping is observed [1]. While detrimental effects from high temperatures are considered in fatigue design standards by empirical formulas (cf. IIW recommendations for welded joints [1]), sub-zero temperatures regarding material selection are completely neglected as long as fracture toughness or Charpy impact toughness requirements are met [2, 3]; however, recent publications observed increasing fatigue strength at sub-zero temperatures down to the transition temperature to brittle material behaviour [4-10].

In the past, fatigue testing at sub-zero temperatures often focused on either fatigue crack growth rate testing at sub-zero temperatures [6-9, 11-14] or on butt-welded joints for stress-life (S-N) testing [15-21]. There are only few studies on fillet welded joints [5, 21-24], which is surprising because they are often more critical from a fatigue perspective. In a recent study, Braun et al. [5] measured an increase in fatigue strength of approximately 20% for cruciform welded joints at temperatures of -50 °C. Besides the scarcity of fatigue test data of welded joints at sub-zero temperatures, applicability of typical fatigue assessment methods to such scenarios is also missing. In the present study, local fatigue assessment methods are applied to fatigue test data by Braun et al. [5] in order to analyse the effect of sub-zero temperatures on fatigue assessment procedures for welded joints. The aim of the study is to calculate the change in fatigue strength with respect to the prediction accuracy of local stress-based fatigue assessment methods. From past studies differences in accuracy of fatigue assessment methods are known to depend on the assessed failure location and weld type [25, 26]. Thus, two different failure locations are analysed within this study.

The analysed data consists of two types of cruciform joint specimens, i.e. load carrying and non-load-carrying (also referred to as double-sided transversal stiffener), with fillet welds. Load carrying fillet welded joints may either show crack initiation from the weld toe or weld root depending on the design of the weld, see Fig. 1. The governing parameter leading to either form of failure initiation is the ratio between throat thickness and plate thickness $a/t$; sometimes the weld height to thickness ratio $H/t$ is also used [27, 28].

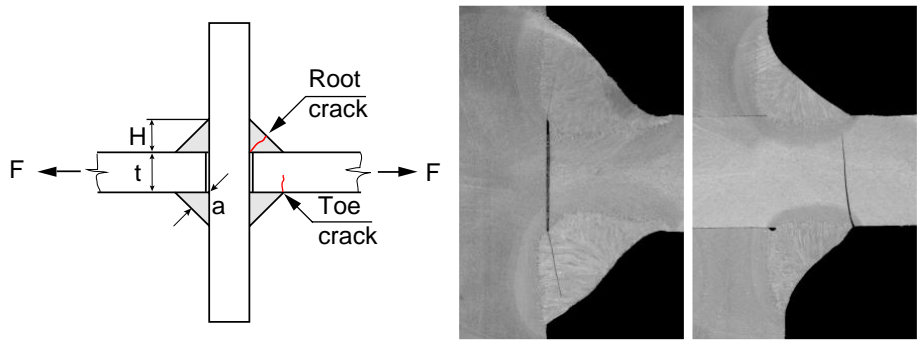

**Fig. 1:** Possible failure locations in fillet welded structures at weld toe and weld root

Nominal and local fatigue assessment methods, suitable for fatigue assessment of both failure initiation sites, are considered within this study. The considered methods can be grouped according to the stress region used for assessment into nominal stress, structural stress, and notch stress approaches. Besides the typical nominal stress assessment, three structural stress methods (structural hot-spot stress, structural stress linearization, and Xiao and Yamada's 1mm stress) and the effective notch stress method are considered. Nominal and structural hot-spot stress approaches are likely the most widely applied methods in fatigue assessment of weld toe failure, while notch stress approaches are frequently applied also in weld root failure assessment, especially in the automotive industry.

An introduction into the considered methods is given in Section 2, then applied in Section 4 to assess the fatigue test results of fillet welded joints at sub-zero temperatures. The fatigue test results are presented in terms of nominal stresses in Section 3. In Section 5 the temperature effect on assessment procedures is discussed and the prediction accuracy of the local fatigue assessment methods is compared for weld toe and weld root failure. Finally, modification factors for fatigue assessment of structures exposed to sub-zero temperatures are presented.

## 2. Background on fatigue assessment methods for fillet-welded joints

Several methods are applicable to fatigue assessment of weld toe and weld root failure of fillet welded joints, usually grouped into global (nominal stress) and local methods (structural and notch stress). Within the nominal stress method, any local stress-raising effects from structural details or welds are neglected in stress assessment but are indirectly considered in the fatigue classes (FAT class). Structural stress methods, on the other hand, consider effects due to the structural geometry but neglect the local stress increase due to the weld. This local stress increase in the vicinity of a weld is only considered in notch stress methods.

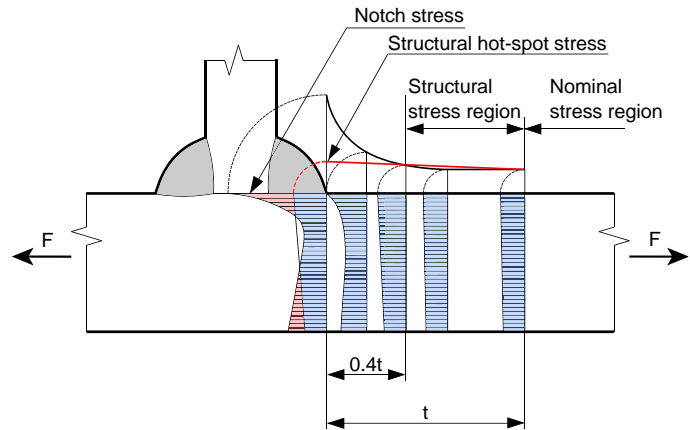

**Fig. 2:** Local stress increase in the vicinity of welded joints with corresponding through-thickness and surface stress profile and fatigue stress regions adopted from [29]

### 2.1. Structural stress methods

The foundation of local methods in general was laid in the 1960s by Haibach [30]. In order to measure the local structural stress ($\sigma_s$) increase due to the structural configuration, he performed strain gauge measurements of the structural strain 2 mm from the weld toe. Thereby, he showed how the increase of the stress in the vicinity of weld toes can be used to assess fatigue strength of different weld details by a master curve. Different approaches have been suggested over the years, which can be grouped into three different categories for avoiding the influence of the nonlinear stress increase at the weld toe [29]. The three approaches, and one corresponding method for each type, are presented in Fig. 3 (a) to (c) beside the effective notch stress method that will be introduced in section 2.2. The three types of structural stress $\sigma_s$ approaches are:

1. an extrapolation of the stress towards the local notch in the area of almost linear stress increase (Fig. 3 (a)),
2. the linearization of the stress either through the plate thickness or in a section through the fillet weld (Fig. 3 (b)), or
3. extraction of the stress component in a single point in the vicinity of the notch, e.g. 1 mm according to Xiao and Yamada [31, 32] (Fig. 3 (c)).

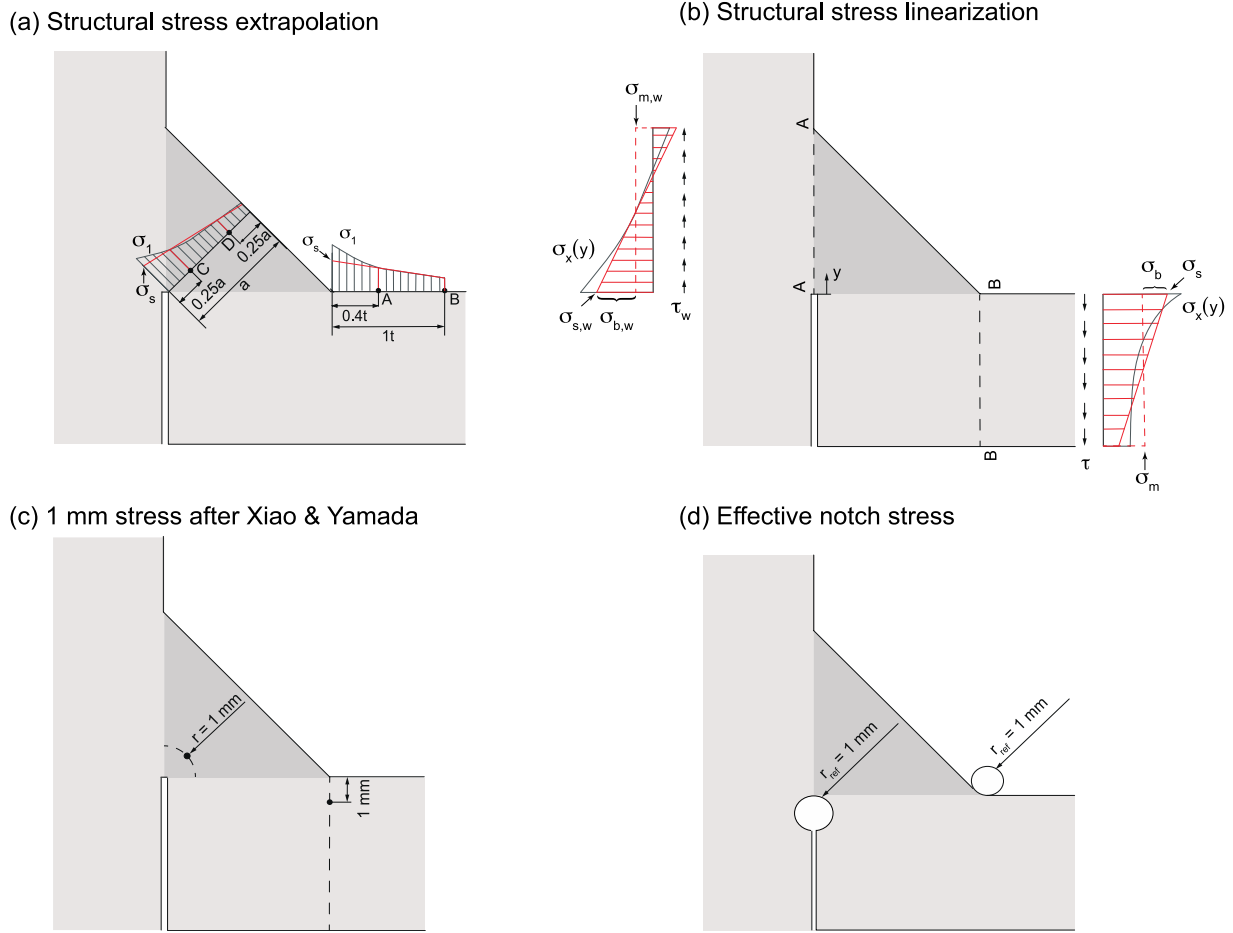

**Fig. 3:** Different local approaches for fatigue assessment illustrated by a single-sided fillet weld

The applicability of different structural stress approaches has been proven by a number of comparative studies for non-load carrying [25, 26] and load-carrying fillet welded structures [33, 34], and of round-robin studies [35, 36]. Summaries of the study results and recommendations on how to apply the methods can be found in Niemi et al. and Fricke [29, 37] for fatigue assessment of weld toe and in Fricke [28] of weld root failure. Below, the details of the three different structural stress methods considered in this study are presented.

### 2.1.1. Structural stress extrapolation approach

Probably the mostly applied structural stress method for weld toe assessment is the surface extrapolation of the structural stress $\sigma_s$ component from reference points (A and B in Fig. 3(a)) that are related to the plate thickness $t$, leading to the so-called structural hot-spot stress. This extrapolation is usually performed by extraction of the first principal stress $\sigma_1$ at 0.4 and 1.0 times the plate thickness. In special cases, other types of extrapolation are recommended; further information can be found in Niemi et al. [29]. This assessment method is associated with the FAT100 master curve for weld toe failure of non-load carrying fillet welds, while FAT90 is recommended for load-carrying fillet welds [1]. Later, the idea of an extrapolation of the structural hot-spot stress was extended to weld root failure by Sørensen et al. [38]. They recommended an extrapolation from the quarter points (C and D in Fig. 3(a)) along the fillet weld bisector. They proposed different mesh subdivisions, with either linear, quadratic brick, or quadratic tetrahedron elements. Linear tetrahedron elements are not to be applied. In this study, the version based on four elements along each line is applied due to its simplicity in preparing the model. For first principal stresses, FAT57 class was proposed in conjunction with a fixed slope exponent $k = 3$ of the S-N curve.

### 2.1.2. Structural stress linearization approach

Another approach is based on the idea of structural stress linearization through the thickness of the base plate (section B-B in Fig. 3(b)) or the fillet weld. Radaj [39] summarized different approaches and showed that the through thickness linearization yields roughly the same result for weld toe failure as the surface extrapolation. This idea was later extended by Fricke et al. [40] to weld root fatigue. High bending stress portion can be assessed as in the case of single-sided fillet welds [40]. The structural weld stress $\sigma_{s,w}$ (sum of membrane $\sigma_{m,w}$ and bending stress $\sigma_{b,w}$ portion) can be obtained by linearization of the stress normal to the leg section $\sigma_x(y)$ as presented in Fig. 3(b) from Eq. 1 to 3:

$$\sigma_{s,w} = \sigma_{m,w} + \sigma_{b,w} \tag{1}$$

$$\sigma_{m,w} = \frac{1}{l} \int_0^l \sigma_x(y)\, dy \tag{2}$$

$$\sigma_{b,w} = \frac{1}{l^2} \int_0^l \sigma_x \left(\frac{l}{2} - y\right) dy \tag{3}$$

with $y$ as coordinate along the leg length $l$ (section A-A in Fig. 3(b)). The structural stress $\sigma_s$, as well as shear stresses $\tau$ in the base plate or $\tau_w$ in the fillet weld, can be linearized in the same way. It should be noted that the approach using the stresses in the weld leg is limited to relatively small shear stresses (up to about 20% of the normal stress), otherwise the crack path may differ considerably from the leg section; structural stress has then to be computed in a different section [41].

For coarse meshes it is generally recommended to use nodal forces instead of nodal or element stresses, otherwise the acting stress can be underestimated by up to 50% [34]. Moreover, due to the singularity, no stress convergence can be achieved based on nodal stress results. For first principal stresses, FAT80 class in conjunction with a fixed slope exponent $k = 3$ was proposed for the linearization through the leg section [41].

One problem of the stress linearization and extrapolation arises for thick plates. Size effects due to large plate thickness can only be included by means of a thickness correction function [42]. One possibility for considering thickness effects directly is using a structural stress approach of the third group, i.e. the assessment of the structural stress in a single point.

### 2.1.3. Xiao and Yamada's 1 mm stress approach

The most-applied method of this group is the 1 mm stress approach by Xiao and Yamada [31, 32]. They argued that the non-linear stress increase decays 1 mm from the local stress raiser, i.e. weld toe or weld root. It is recommended to use the proposed fatigue design curves with the stress normal to the expected crack path at 1 mm below the surface. For weld toe failure, this is the stress component parallel to the direction of applied loading. For weld root failure, an assessment along a quarter circle with radius $r = 1$ mm in 3° steps is proposed to find the location of maximum normal stress (i.e. the tangential stress $\sigma_\theta$ in polar coordinates). The limitation of the 1 mm stress approach are thin welded joints due to the steep gradient in through thickness direction [43]; however, it is well suited for thicker plates ($t > 8$ mm). In such cases, elements with quadratic shape function as large as 0.5 mm are sufficient for assessment at weld toes [26]. For weld toe failure assessment, FAT100 class is recommended again and FAT65 is recommended for weld root failure. While this method has been successfully applied to weld toe failure, it is not often used for cruciform joints failing from the weld root.

### 2.2. Effective notch stress approach

Since all structural stress approaches focus on the almost linear stress increase between the far-field stress and the non-linear local stress field, local stress effects are covered in the fatigue design curves. This inevitably leads to the necessity of different fatigue design curves for weld toe and root failure due to large differences in local stress level. In order to include the local stress field in the assessment, the effective notch stress approach is one viable alternative.

In welded structures, small notch radii usually occur at the weld transitions. The effective notch stress concept, according to Radaj [39], is based on the assumption of vanishing notch radii ($r_{real} \to 0$ mm). Based on von Mises stress hypothesis, a factor $s = 2.5$ for stress multiaxiality and plane strain condition, and a microstructural support length of $\rho^* = 0.4$ mm (assuming a cast iron-like microstructure in the heat affected zone [44]), an enlargement of the real notch radius $r_{real}$ to a reference radius $r_{ref} = 1$ mm was proposed.

$$r_{ref} = r_{real} + s\rho^* \quad (4)$$

Problems related to the effective notch stress approach, especially for small fillet weld sizes, occur where the enlarged radius at the weld root significantly reduces throat thickness [37]. This problem can be overcome by using smaller reference radii (e.g. 0.05 mm or 0.3 mm) in conjunction with corresponding fatigue design curves or by moving the centre of the fictious radius to maintain the correct cross-sectional area. A review on different notch stress approaches can be found in Baumgartner [44]. In conjunction with the 1 mm radius, FAT225 class is recommended for weld toe and root failure assessment. In the following section, the fatigue test data of fillet welded joints is introduced; the aforementioned local approaches will be applied to this.

## 3. Fatigue test results at sub-zero temperatures

In Braun et al. [5], fatigue test results of a load-carrying cruciform joint with fillet welds and double-sided transversal stiffener, leading to weld root and weld toe failure respectively, are presented for a load ratio $R = 0$ in the temperature range from 20 °C down to -50 °C. This temperature range is representative for a ship traveling through Arctic regions year round [45]. The specimens were produced from two structural steels: a normalized mild steel (S235J2+N) and a fine grained thermomechanically-rolled steel (S500G1+M). Before the tests, the misalignment and local weld geometry of every specimen was measured based on the curvature method [46, 47].

### 3.1. Nominal stress results

For weld toe fatigue nominal stresses can be derived directly from the nominal stress acting in the adjacent plate. When assessing weld root fatigue, however, the nominal stress is based on averaged stress components in the weld throat that are also used for static design of welds [28]. In this context, the definition of throat thickness based on Eurocode 3 [48] is applied, i.e. fitting the largest possible triangle into a fillet weld. Thereby, no load transfer is assumed through excessive weld overfill. The validity of this assumption seems reasonable, since almost the same fatigue strength is derived for both steel strengths if plotted over applied force instead of stress (the fillet welds of the S500 specimens were more convex). The throat thickness, according to Eurocode 3, can be calculated from the leg lengths ($z_1$ and $z_2$) with

$$a = (z_1 + z_2)/\sqrt{z_1^2 + z_2^2} \qquad (5)$$

The effect of temperature on fatigue strength is highlighted by presenting all results relative to its nominal FAT class (FAT80 for transversal stiffener and FAT36 for cruciform joints failing from the weld root) in Fig. 4(a). Moreover, in order to calculate deviation between experimental and predicted number of cycles to failure—based on fatigue design curves mentioned (probability of survival of 97.5%)—and to be able to compare the accuracy with local fatigue assessment methods, the results are presented in a deviation plot in Fig. 4(b). For the following comparison of local fatigue assessment methods, the same symbols and colours are used to ease differentiation between the test series. All test series of cruciform joints with weld root failure are denoted with a "C" and the transversal stiffener series with a "T". Moreover, filled symbols are used for S500 steel and empty for S235 steel. The colour scheme for the temperature range is black (room temperature), red (-20 °C), and blue (-50 °C).

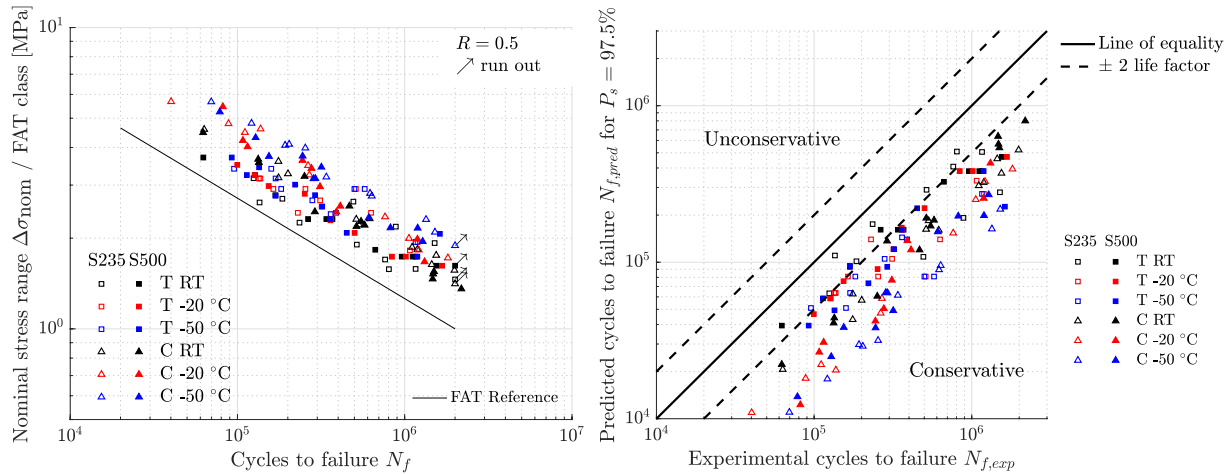

**Fig. 4:** Nominal stress results of S235 and S500 cruciform joint ("C" symbol) and transversal stiffener specimens ("T" symbol) normalized by corresponding FAT36 and FAT80 weld detail classes; corrected to $R = 0.5$

All results are corrected to a stress ratio of $R = 0.5$ based on the mean stress correction factors calculated by Braun et al. [5] to allow a comparison with fatigue design curves (1.02 and 1.08 for S235 and S500, respectively, and both joint types). All cruciform joints failed from the weld root, in particular the results for sub-zero temperatures lie far above the corresponding design curve FAT36. For cruciform joints failing from the weld root, axial misalignment $e$ and residual stresses are well known to govern fatigue strength [49, 50]. Comparing the axial misalignment of the test data, presented here, with data by Andrews [49], it becomes clear that the FAT class for weld root failure of cruciform joints is reached only in cases when the axial misalignment approaches the magnitude of plate thickness ($e/t = 1$). For $e/t = 1$, the fatigue strength in Andrews [49] was about half the strength for $e/t = 0.25$. Although misalignment levels are small for the presented test data (weld quality fulfils for most specimens the requirements for class B of ISO5817:2014), misalignment effects have a significant influence on fatigue strength and on the derived fatigue design curves [47]; these are considered for all fatigue assessment methods according to the following procedure.

### 3.2. Consideration of misalignment effects

Due to welding and the unsymmetrical cooling process, welded plates show varying axial (offset $e$) and angular misalignment (angle $\alpha$). It is well known that misalignment leads to high additional bending stresses at the welds; this significantly reduces fatigue life. In order to differentiate the misalignment effect from the temperature effect, the test results are corrected for their misalignment-induced secondary bending stresses. For this purpose, the applied nominal stress ranges are corrected according to IIW recommendations [1] for their actual angular and axial misalignment by superposition of stress magnification factors for axial $k_{m,e}$ and angular misalignment $k_{m,a}$ with

$$k_m = 1 + (k_{m,e} - 1) + (k_{m,a} - 1) \tag{6}$$

Berge and Myhre [52] and Maddox [53] were the first to publish equations for the calculation of stress magnification factors for weld toe fatigue ("WT" index in Eq. 7 and 8) based on beam theory. For fixed ends, same plate thickness $t$, and different plate lengths ($l_1$, $l_2$) the formulas are

$$k_{m,e,WT} = 1 + \frac{\lambda \cdot e \cdot l_1}{t(l_1 + l_2)}, \text{ with } \lambda = 3 \text{ for fixed ends} \tag{7}$$

$$k_{m,a,WT} = 1 + \lambda \cdot \alpha \cdot \frac{l_1 \cdot l_2}{t(l_1 + l_2)}, \text{ with } \lambda = 3 \text{ for fixed ends} \tag{8}$$

In the 1990s, Andrews [49] published an equation to calculate the stress magnification due to axial misalignment for weld root fatigue in load-carrying cruciform joints $k_{m,e,WR}$.

$$k_{m,e,WR} = 1 + \frac{e}{t + H} \tag{9}$$

Since a formula for the assessment of angular misalignment effect on weld root fatigue has not yet been developed, it is assumed that the secondary bending effect ratio caused by axial and angular misalignment between weld toe and root is approximately constant.

$$\frac{k_{m,a,WR}}{k_{m,a,WT}} = \frac{k_{m,e,WR}}{k_{m,e,WT}} \qquad (10)$$

In the *nominal stress approach*, a certain degree of misalignment is already covered in the fatigue design curves. Thus, the nominal stress results are only corrected to an effective stress magnification factor if the misalignment effect exceeds the included misalignment levels given in the IIW recommendations [1, 29]. For this purpose, stress magnification derived from the measured misalignment prior to the test shall be corrected with the magnification included in the FAT classes (e.g. 1.25 for transversal stiffener); however, for weld root failure, no information is given in the IIW recommendations. It is assumed that the level is higher than for transversal stiffener, since a higher level of misalignment is included for weld toe failure of cruciform joints (1.45). In general, the measured angular misalignment ($\alpha$) and the axial misalignment ($e$) was generally low, see Fig. 5. Here, the measured mean and standard deviation (STD) misalignment values and stress magnification levels are listed for all specimen types. As can be seen from Fig. 5, the stress magnification factor of all transversal stiffener is below the level given by the IIW recommendations. Thus, no misalignment correction is applied for the *nominal stress approach*.

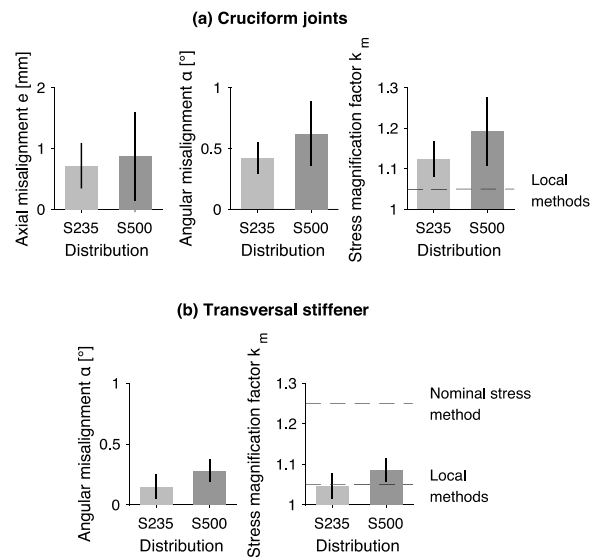

**Fig. 5:** Mean and standard deviation of measured misalignment and calculated stress magnification levels of cruciform joint (a) and transversal stiffener (b) specimens

For the *local approaches*, only a small constant amount of misalignment (1.05) is covered in the structural hot spot and effective notch stress design curves, whereas no information is available on the misalignment level included in the Xiao and Yamada design curve [32], assuming here to be negligible. As the misalignment effects in the tests are larger than 5%, they have been included in the local stress analysis. Moreover, for the sake of simplicity and to be consistent the actual stress magnification has been considered of all specimens also below 5% and no correction for already included misalignment is applied, since this information is not available for all *local approaches*.

4. **Fatigue assessment based on structural stress methods and effective notch stress method**

For all local fatigue assessment methods, quarter FE models of the welded joints are built according to IIW or other recommendations for the corresponding methods [28, 29, 31, 37, 38]—or for the 1 mm stress approach by Xiao and Yamada, according to the recommendations by the authors for weld root assessment [32]. For weld toe assessment, a coarser mesh with global element size of 0.1 mm is used. The meshes for weld root fatigue assessment are presented in Fig. 6. Since the deviation in weld throat thickness is small for both steel types (about 0.2 mm standard deviation) the mean throat thickness is used for modelling (i.e. $a$ = 5.71 mm and $a$ = 5.85 mm for S235 and S500 cruciform joints, respectively).

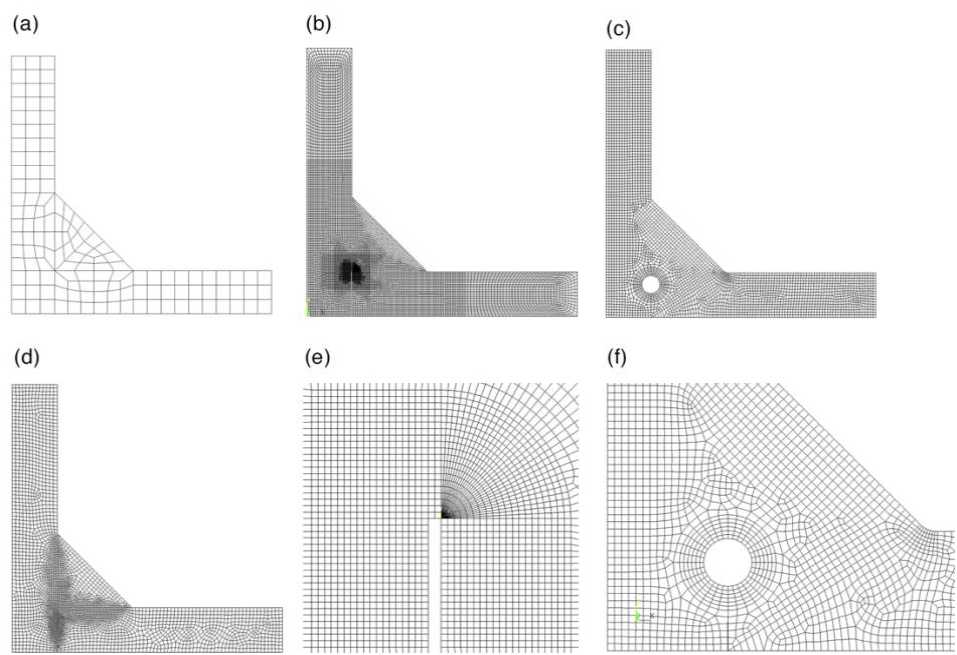

**Fig. 6:** FE meshes of the cruciform joint specimen with weld root failure used for the different fatigue assessment methods: (a) structural stress extrapolation, (b) structural stress linearization, (c, d) Xiao and Yamada 1 mm stress concept, and (e, f) effective notch stress concept

In this study, the version based on four elements along each line is applied for the structural stress extrapolation through the weld throat. For the stress linearization along the leg length, 50 elements are used and the model for the 1 mm stress concept is built according to the recommendations by Xiao and Yamada (one element every 3°, slit of 0.1 mm and a mapped mesh), see Fig. 6(d) and [32] for further details. The notch stress model presented in Fig. 6 (e) and (f) is built according to IIW recommendations [37], with 32 elements over 360°. Moreover, for all methods, 8-node plane strain elements (plane183 in Ansys) are used. The secondary bending stress due to misalignment is applied in a simplified way as an increased nominal stress, which is thought to be conservative [47].

### 4.1. Fatigue assessment results

By applying a unit stress to the FE models presented in Fig. 6, stress concentration factors for structural stress ($K_S$) and notch stress ($K_t$) methods are calculated. The calculated stress concentration factors and corresponding fatigue design curves (given by the characteristic fatigue strength FAT) are listed in Tab. 2 and Tab. 3. The stress result of each specimen is thus calculated by multiplying the nominal stress (magnified by misalignment effects) with the corresponding stress concentration factor for each method.

**Tab. 2:** Summary of calculated stress concentration factors for structural stress ($K_S$) and notch stress ($K_t$) for weld toe (transversal stiffener) failure with corresponding fatigue strength ($P_s \approx 97.5\%$, $N = 2 \cdot 10^6$)

| Assessment method | Characteristic fatigue strength | Stress concentration factor for structural stress ($K_S$) and notch stress ($K_t$) |
| --- | --- | --- |
| Stress extrapolation | FAT100 | 0.999 |
| Stress linearization | FAT100 | 0.996 |
| 1mm concept | FAT100 | 1.055 |
| Effective notch stress | FAT225 | 2.417 |

**Tab. 2:** Summary of calculated stress concentration factors for structural stress ($K_S$) and notch stress ($K_t$) for weld root (cruciform joints) failure with corresponding fatigue strength ($P_s \approx 97.5\%$, $N = 2 \cdot 10^6$)

| Assessment method | Characteristic fatigue strength | Stress concentration factor for structural stress ($K_S$) and notch stress ($K_t$) |
| --- | --- | --- |

|  |  | $a = 5.71$ mm | $a = 5.85$ mm |
|---|---|---|---|
| Stress extrapolation | FAT57 | 1.327 | 1.294 |
| Stress linearization in leg section | FAT80 | 1.610 | 1.585 |
| 1mm concept | FAT65 | 1.201 | 1.184 |
| Effective notch stress | FAT225 | 4.006 | 3.941 |

The results for the four analysed fatigue assessment methods are presented in Fig. 7 to 10 with the corresponding fatigue design curves and the deviation between experimental and predicted number of cycles to failure for a probability of survival of approximately 97.5% according to Eq. 11. It is not known if the design curves of all methods are derived by using a 95% confidence interval. Some might be derived based on the mean S-N curves minus two standard deviations; however, it is generally assumed that the difference between both approaches is negligible [1].

$$\text{dev} = \log N_{f,exp} - \log N_{f,pred,97.5\%} \tag{11}$$

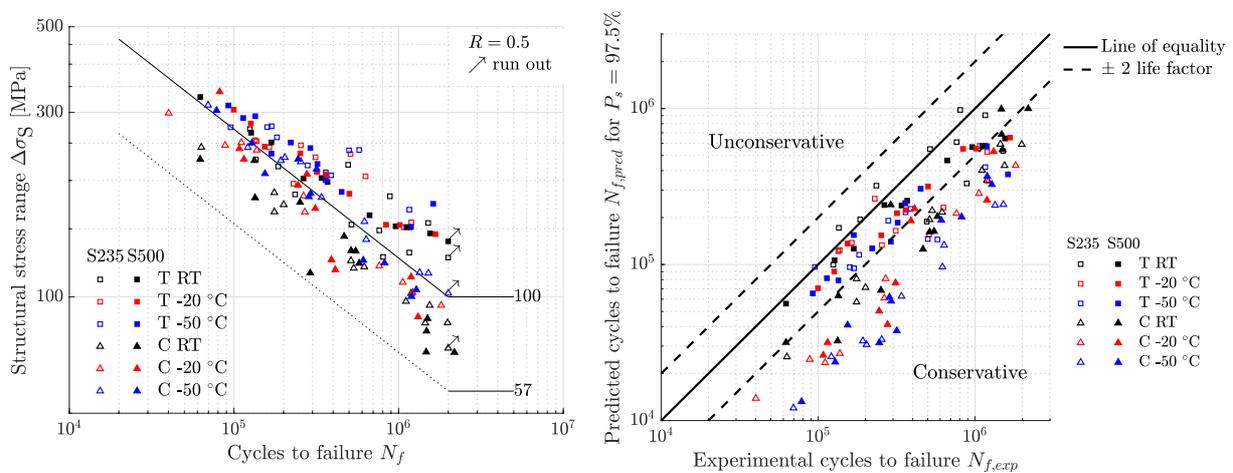

**Fig. 7:** Structural stress extrapolation results converted to $R = 0.5$ with corresponding FAT classes for transversal stiffener ("T", FAT100) and weld root failure of cruciform joints ("C", FAT57)

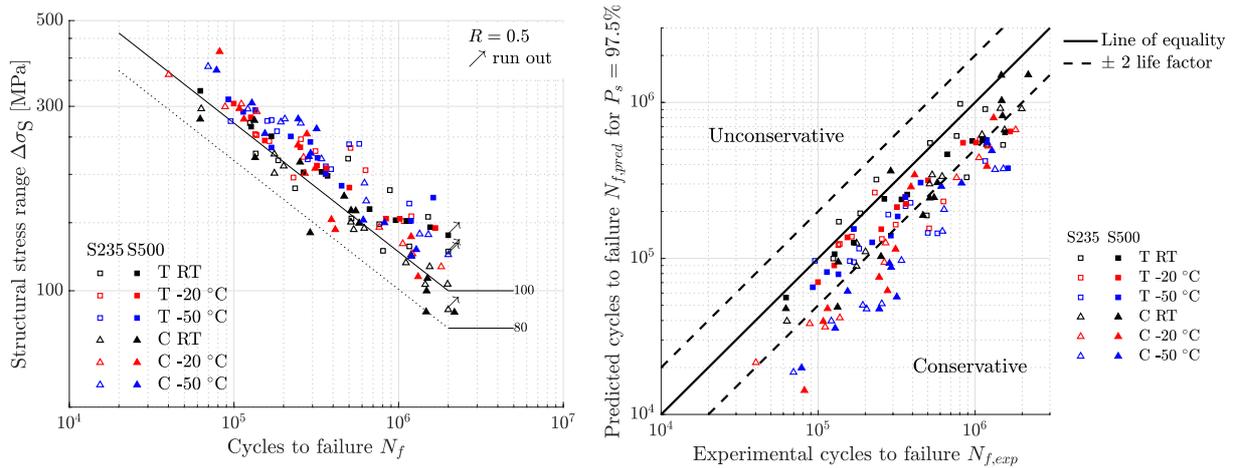

**Fig. 8:** Structural stress results based on stress linearization converted to $R = 0.5$ with corresponding FAT classes for transversal stiffener ("T", FAT100) and weld root failure of cruciform joints ("C", FAT80)

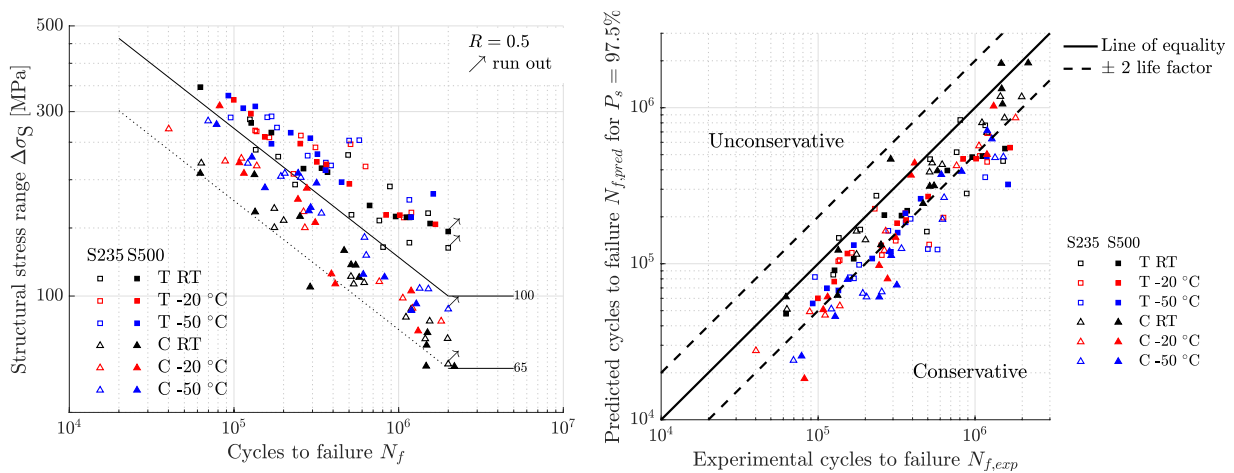

**Fig. 9:** Structural stress results of the 1 mm concept by Xiao and Yamada [31, 32] converted to $R = 0.5$ with corresponding FAT classes for transversal stiffener ("T", FAT100) and weld root failure of cruciform joints ("C", FAT65)

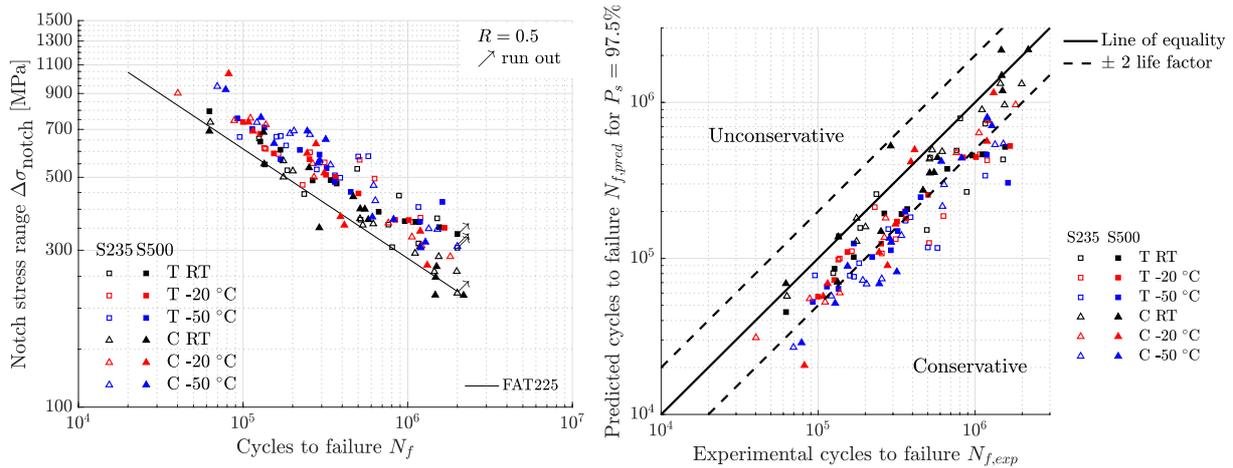

**Fig. 10:** Effective notch stress results converted to *R* = 0.5 with corresponding FAT225 class

Comparing the structural hot-spot and linearization fatigue assessment methods, a similar trend is observed with the results for the cruciform joints being assessed more conservatively than the results of the double-sided transversal stiffener. On the other hand, the test results of the cruciform joints with weld root failure are partially below the corresponding design curves of Xiao and Yamada and effective notch stress concepts at room-temperature. Moreover, a number of transversal stiffener specimens lie below the corresponding design curves for all structural stress methods, but only one specimen lies below the FAT225 curve of the effective notch stress method. In general, the results for all four methods deviate less from the design curves than for the nominal stress approach.

## 5. Discussion

### 5.1. Temperature effect on assessment procedures

In order to ease comparison, the distribution of the deviation for the four analysed local fatigue assessment methods is presented as probability density functions with underlying histograms in Fig. 11. Due to the aforementioned large deviation of load-carrying cruciform joint results for weld root failure from the corresponding nominal design class (FAT36), the results for weld toe and weld root failure are evaluated separately.

First, the results for room temperature fit well a normal distribution; this confirms the general expectation of lognormal distributed fatigue test results of welded joints. The results for lower temperatures do not seem to fit a normal distribution as well; however, the applicability of a normal distribution to describe the deviation is confirmed by Chi-square goodness-of-fit testing. The null hypothesis of normal distributed logarithmic deviations is confirmed for all six data sets based on a typical significance level of 5%. It can thus be assumed that the fatigue test results themselves are lognormally distributed.

The apparent difference between the histogram and the distribution function may be related to the size of the dataset, which consists of four test series for each temperature, thus only two series (20-25 specimens) for each failure location. Interestingly, the standard deviation of the deviation data for the specimens with weld toe failure changes only slightly with temperature, although it increases as temperature decreases for the specimens with weld root failure. A change of the slope parameter of the crack growth curve, which is frequently reported for fatigue test at sub-zero temperatures close to the ductile-brittle transition temperature of the material [6, 7, 9, 14], may be a reason for this behaviour. In Braun et al. [4] Charpy notch impact test results for the same material and welding process are reported with a ductile-brittle transition temperature (27J criterion) of the weld metal at approximately -5 °C and -40 °C for the S235J2+N and S500G1+M steel respectively. The transition temperature of the base material and in the heat-affected zone was, for both materials, below the lowest test temperature of the fatigue test results in this study (-50 °C). Consequently, it is likely that the cruciform joints with cracks in the weld material were in, or below, the fatigue transition regime at that test temperature. The slope of the S-N curve is therefore affected by a changed crack growth slope exponent. The comparison with a fatigue design curve based on a fixed slope exponent of $k = 3$ leads therefore to an increased standard deviation of the data deviation at sub-zero temperatures than for room temperature. Moreover, the fracture surfaces presented in Braun et al. [5] support this conclusion. Finally, significant differences in prediction accuracy between the different methods are apparent in Fig. 11.

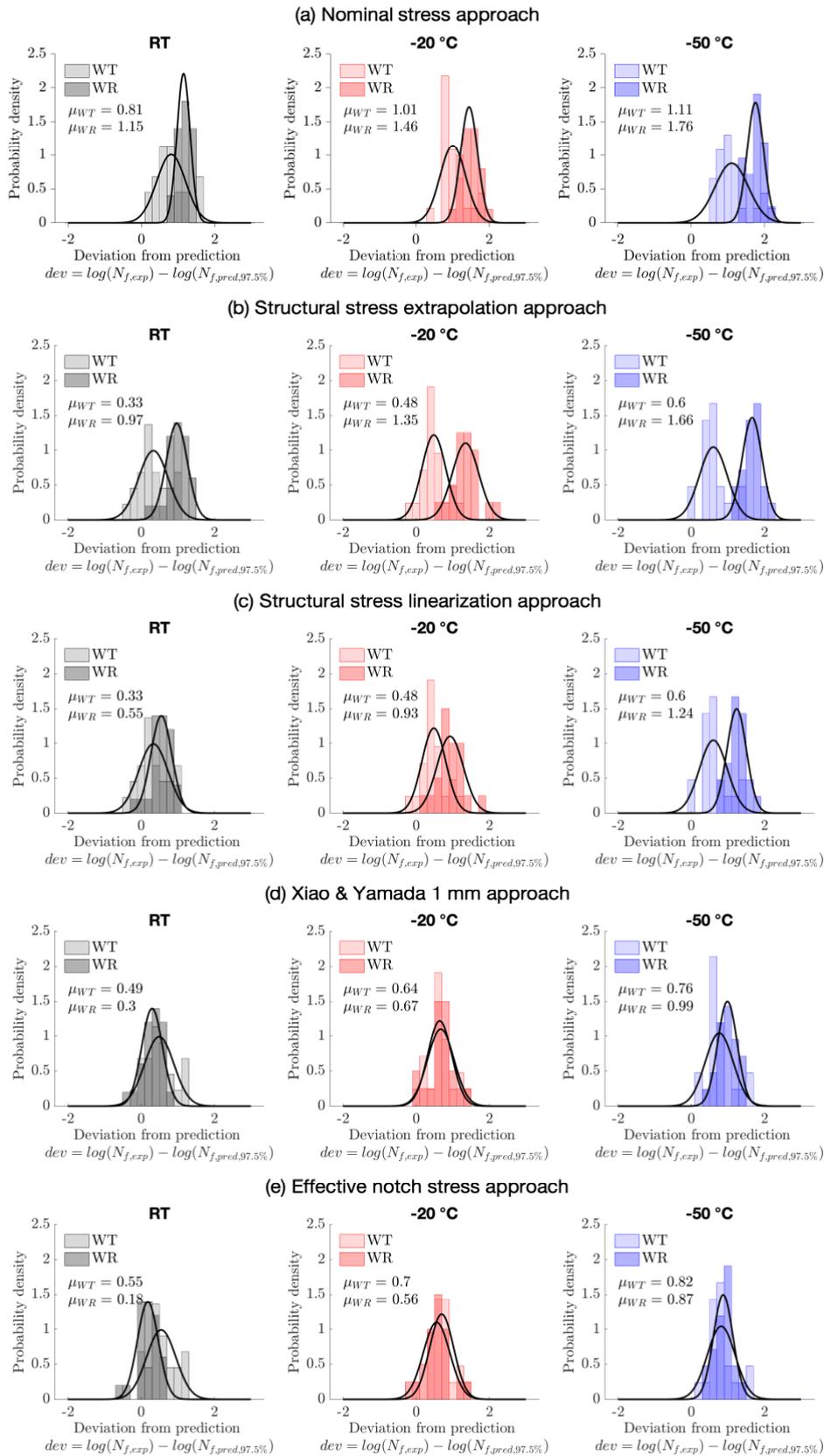

**Fig. 11:** Distribution of relative deviation between experimental and predicted number of cycles to failure ($P_s \approx 97.5\ \%$, $N = 2 \cdot 10^6$) for (a) nominal stress, (b) structural stress extrapolation, (c) structural stress linearization, (d) Xiao and Yamada 1 mm approach, and (e) effective notch stress approach separately for weld toe and weld root failure

## 5.2. Comparison of local fatigue assessment approaches for sub-zero temperatures

The mean deviation from prediction (97.5% probability of survival) is presented in Fig. 12, including scatter bands for the standard deviation (SD) of the deviation results. As can be seen from Fig. 12, the prediction accuracy is quite different for both failure locations and methods. First of all, all methods fit well for weld toe failure, but show partially large deviations for weld root failure. In general, the deviation is highest for the nominal stress approach. The structural stress extrapolation and linearization fit well for weld toe failure; however, there is a large deviation between test results and prediction for weld root failure. Moreover, the effective notch stress method and Xiao and Yamada's 1 mm stress approach yield the least over-conservatism for weld root failure. While a small number of specimens tested at room temperature lie below the corresponding design curves of all local fatigue assessment methods, none lie below the nominal stress methods design curves.

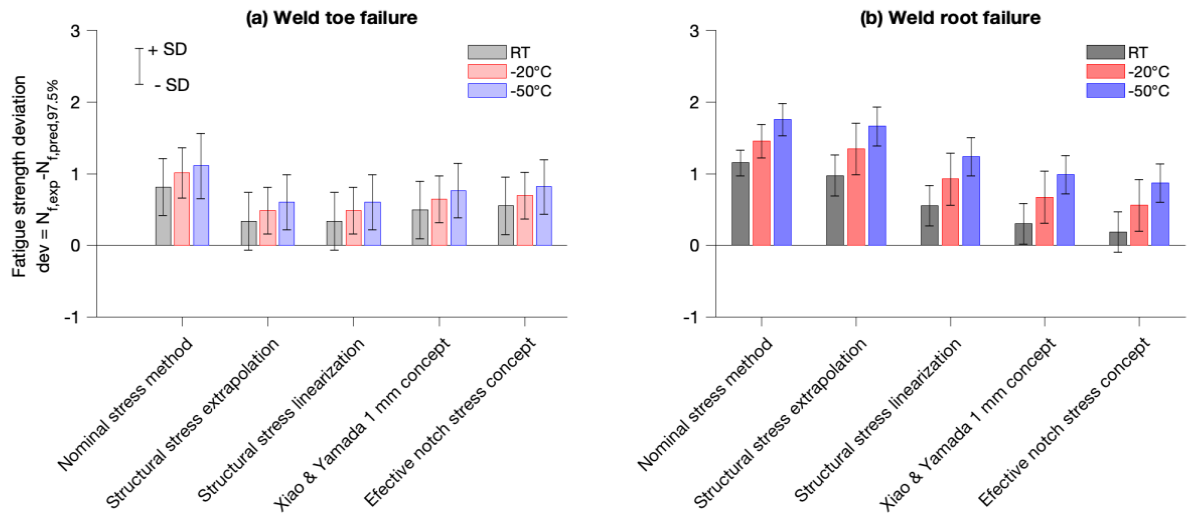

**Fig. 12:** Deviation between experimental an estimated fatigue strength ($P_s \approx 97.5\,\%$, $N = 2 \cdot 10^6$) for (a) weld toe failure and (b) weld root failure; corrected to $R = 0.5$

The results presented in Fig. 12 can be used as modification factors for fatigue assessment at sub-zero temperatures in conjunction with Eq. 11; however, more fatigue test data is required to verify the observed fatigue strength increase at sub-zero temperatures. The large deviation between experimental and predicted fatigue strength for weld root failure by means of nominal stress and structural stress extrapolation methods leads to the question of the applicability of the underlying fatigue design curves if misalignment is small or considered in fatigue assessment [47]. The results of the cruciform joint match the design curve for the effective stress method the best, which is thought to be related to the small misalignment level of the specimens (T-joints loaded in bending) that were used to derive the design curve. The results for weld toe failure, however, show much less deviation between the different methods, which would permit general modification factors to be applied for the nominal stress and all local assessment methods for sub-zero temperatures.

## 6. Conclusions

This study investigated the applicability of local fatigue assessment methods for welded joints in engineering structures exposed to sub-zero temperatures. For this purpose, fatigue test results by Braun et al. [5] of two fillet weld details with weld toe and weld root failure made of two structural steels in the temperature range of 20 °C down to -50 °C were evaluated using different structural- and notch stress methods found in international standards and literature. The results are compared to the nominal stress approach. From the investigation, the following conclusions are drawn:

- The increase of fatigue strength at sub-zero temperatures requires modification factors for all types of fatigue assessment.

- Large differences in prediction accuracy are found for weld toe and weld root failure. All concepts are leading to conservative assessment results at sub-zero temperatures; however, for some methods and especially for weld root failure the results are very conservative.

- In general, the higher deviation between experimental and predicted fatigue strength for weld root failure is thought to be related to the low level of misalignment of the test specimens in this study and not considered to be caused by a detrimental effect of residual stresses. Thus, a large deviation is observed for all fatigue assessment methods of weld root failure, except for Xiao and Yamada's 1 mm stress method and the effective stress method.

- The standard deviation of the logarithmic deviation between the predicted and experimental fatigue life is almost constant for all three test temperatures of weld toe failure (crack growing through the heat-affected zone and base material) and is increasing for weld root failure (cracks in the weld metal). This effect is thought to be related to the change of fatigue crack growth slope exponent around the fatigue ductile-brittle transition temperature. For the weld metal, transition temperatures based on Charpy impact testing were reported in another study [4] to be at approximately -5 °C and -40 °C for the S235J2+N and S500G1+M steel, respectively.

- The assumption of lognormal-distributed fatigue test results of welded joints is verified by means of Chi-square goodness-of-fit testing for room temperature and sub-zero temperature fatigue test results.

- For weld toe failure nominal stress as well as structural stress extrapolation and linearization approaches yield the best results. For weld root failure, the highest accuracy is achieved for the effective notch stress approach and Xiao and Yamada's 1 mm stress concept.
- Xiao and Yamada's 1 mm stress concept has rarely been applied to weld root assessment in previous studies; the results, however, clearly show that the prediction accuracy is comparable to the effective notch stress method.
- Finally, all local methods—as well as the nominal stress approach—are applicable to sub-zero temperatures; further research is required to establish modification factors for sub-zero temperature fatigue assessment.

**Acknowledgements**

The work was performed within the research project ESM-50 Fatigue of welded structures at sub-zero temperature, funded by the German Research Association of the Working Group of the Iron- and Metal-processing Industry e.V. as part of the Donors' Association for the Promotion of Sciences and Humanities in Germany under project number AVIF-No. A 301. The project was accompanied by a working group of the German Shipbuilding and Ocean Industries Association e.V. (VSM) / Center of Maritime Technologies e.V. (CMT). The authors would like to thank Robert Scheffer for the contribution of figures to this study, Shi Song for performing the geometry measurements, and Rachael Elizabeth Wu for proof-reading the manuscript.